\documentclass[onecolumn]{elsarticle}
\usepackage{graphicx}
\usepackage{amsmath}
\usepackage{amssymb}
\journal{Physics Letters B}

\begin{document}
\begin{frontmatter}
%
%
\title{Short Distance Freedom of Quantum Gravity}
%
\author{Gaurav Narain\footnote{Electronic address: {\tt gaunarain} at {\tt imsc.res.in}},\,
and Ramesh Anishetty\footnote{Electronic address: {\tt ramesha} at {\tt imsc.res.in}}
\\
~\\
The Institute of Mathematical Sciences, Taramani, Chennai 600113, India}
%
\begin{abstract}
Fourth order derivative gravity in 3+1-dimensions is 
perturbatively renormalizable and is shown to describe a unitary theory of
gravitons in a limited coupling parameter space. 
The running gravitational constant 
which includes graviton contribution is computed. 
Generically, gravitational Newton's constant vanishes at short 
distances in this perturbatively renormalizable 
and unitary theory. 
\end{abstract}

\begin{keyword}
Renormalization Group \sep Higher Derivative Gravity \sep Unitarity
\end{keyword}

\end{frontmatter}
%

\section{Introduction and Theory}
\label{intro}

Newton's gravitational constant $G$ in classical Einstein-Hilbert (EH)
action is one of the smallest 
fundamental coupling constant with immense success in 
Cosmology \cite{Dodelson, Weinbergbook}. Yet it is less understood
quantitatively when we quantize the theory. The primary reason being
that in $3+1$ space-time dimensions this coupling constant behaves as
$(Mass)^{-2}$ and the theory has ultra-violet (UV) divergences. 
It was known \cite{Utiyama1962} that even quantum matter fields in the presence of 
background gravity demands that up to fourth order metric derivative terms
need to be considered for UV regularization. Renormalization
group (RG) studies of EH action \cite{Stelle} also 
suggest that we augment the gravitational 
action with fourth order derivative terms and witness 
UV renormalizability. Most general action functional 
up to fourth order derivative of metric $g_{\mu\nu}$ 
is given by $S$,
\begin{equation}
\label{eq:hdgact}
S
= \int \frac{{\rm d}^4x \sqrt{-g}}{16 \pi G} \left[
2 \Lambda -R + \frac{\omega R^2}{6 M^2}
- \frac{R_{\mu\nu}R^{\mu\nu} - 
\frac{1}{3}R^2}{M^2}
\right] \, ,
\end{equation}
where $R$ is the Ricci scalar and $R_{\mu\nu}$ is the 
Ricci tensor. The last term is also proportional to square 
of conformal Weyl tensor in $3+1$ space-time dimensions, $\omega$
is dimensionless and $M$ has dimension of mass. 
It was first shown in \cite{Stelle} that the action $S$ is UV
renormalizable using the $4-\epsilon$ dimensional regularization
scheme. The primary observation being the action is non-linear in 
metric and hence has many number of basic interactions between 
gravitons. However, in the UV regime the fluctuating metric field (in $3+1$ 
space-time dimension) has zero mass dimension and 
all these terms due to general co-ordinate invariance can 
be tamed by renormalizing the coupling parameters in Eq. (\ref{eq:hdgact}).

In the action $S$, classically if $M^2 \to \infty$ and $\Lambda \to 0$,
only the EH term dominates, thereby accounting for all
the known success of EH action. A deeper look into quantizing 
the action using ADM formalism 
\cite{ADM, boulwarebook, Strominger, Buchbinder} showed that 
indeed $S$ corresponds to the well-known constraints, 
space-time translation and a host of other
constraints. The Hamiltonian itself reduces to a surface
integral just as in pure EH gravity. Furthermore, the Hamiltonian
is shown to be positive if the last term in $S$ is absent and the 
coefficient of $R^2$ term is positive \cite{Strominger, Buchbinder}. 
It is also shown that 
in Minkowskian background metric, the theory describes 
massless spin-2 gravitons, a massive scalar and a
massive spin-2 field. 

Perturbation theory about a flat background 
$g_{\mu\nu} = \eta_{\mu\nu} + h_{\mu\nu}$ 
($\eta_{\mu\nu} = \{1,-1,-1,-1 \}$) was studied in \cite{Salam, Julve}.
Here we investigate  the action $S$ for $\Lambda=0$ in the Landau gauge 
$\partial^{\mu} h_{\rho\mu}=0$.
The Feynman propagator in momentum space is,
\begin{equation}
\label{eq:grav_prop1}
D_{\mu\nu, \alpha\beta} =
 \frac{i \, 16 \pi G}{(2 \pi)^4} \cdot \Biggl[
\frac{(2P_2 - P_s)_{\mu\nu, \alpha\beta}}{q^2 + i \, \epsilon}
+ \frac{(P_s)_{\mu\nu, \alpha\beta}}{q^2 - M^2/\omega + i \epsilon}
- \frac{2 \, (P_2)_{\mu\nu, \alpha\beta}}{q^2 - M^2+ i \epsilon}
\Biggr]
\end{equation}
where $q$ is the momentum of fluctuating field $h_{\mu\nu}$. Various 
spin projectors are  
$(P_2)_{\mu\nu, \alpha \beta} = 
\frac{1}{2} \left[ T_{\mu\alpha} T_{\nu\beta} + 
T_{\mu\beta}T_{\nu\alpha} \right] 
- \frac{1}{3} T_{\mu\nu}T_{\alpha\beta} $, 
$(P_s)_{\mu\nu, \alpha \beta} = 
\frac{1}{3} T_{\mu\nu} \, T_{\alpha \beta}$, 
where $T_{\mu\nu}=\eta_{\mu\nu} - q_{\mu}q_{\nu}/q^2$.
The first term of Eq. (\ref{eq:grav_prop1}) is the massless spin-2
graviton with two degrees of freedom, the second term 
is the scalar (Riccion) of mass $M/\sqrt{\omega}$ and the last
term is that of massive spin-2 with mass $M$ ($M$-mode), which is in 
agreement with the general ADM analysis. Furthermore, the residues
at the pole for graviton and Riccion are positive, while for the $M$-mode
it is negative. The $M$-mode renders the theory 
non-unitary.

We now investigate the issue of the unitarity 
of the higher derivative gravitational action Eq. (\ref{eq:hdgact})
in the context of background Minkowskian perturbation theory. 
The main observation
being that in quantum theory, all the couplings 
cease to be constant and will start running with momentum $\mu$.
Thus effectively at energy scale $\mu$ we have running couplings
$G(\mu)$, $M^2(\mu)$ and $\omega(\mu)$. 
This is an efficient way to incorporate the quantum effects
in the renormalizable theory. In such a context if $M^2(\mu) / \mu^2 > 1$
for all $\mu^2$, then in any scattering process involving gravitons and 
Riccions we will never encounter the $M$-mode particle as an 
intermediate state to any order in Feynman loop 
expansion. Consequently the negative norm 
does not violate unitarity of the scattering matrix in 
graviton and Riccion subspace (physical GR).
To reiterate consider a very high energy graviton scattering
off with another graviton or Riccion. In principal there is 
enough energy to produce an $M$-mode, however the couplings
and the masses are evolving with energy. In the renormalized theory where 
$M^2(\mu^2)/\mu^2 >1$, the mass of $M$-mode is always
greater than the available energy, hence it is never realized on shell.
In a scattering process of high energy graviton and Riccion going to
other gravitons and Riccions with different momenta, $M$-mode 
can give off mass shell contributions, which
affects the real part of the transition amplitude and never to the 
imaginary part (Cutkosky cut) \cite{Diagram}. Since
$M^2(\mu^2)/\mu^2 >1$ implies it cannot go on-shell.
Indeed we can reinterpret $M$-mode as an effective ghost 
normalizing the functional integral as conjectured in
\cite{Salam}. We will show that this can 
be explicitly realized in a suitably defined parameter space of
the action $S$. 

\section{Renormalization Group Analysis}
\label{RGanalysis}

We first make the observation that our choice of parameterization in
Eq. (\ref{eq:hdgact}) naturally shows that the Feynman loop expansion
is the same as perturbation theory in small $G$, namely $G$ 
effectively plays the same role as $\hbar$. To one-loop the beta
function of couplings have been computed using Schwinger-Dewitt 
technique \cite{Dewitt}. Using dimensional regularization in $4-\epsilon$ dimensions
and minimal subtraction \cite{Fradkin,avramidi} 
in the background Landau gauge,
the RG flows are the following \cite{avramidibook}:
\begin{eqnarray}
\label{eq:betaM2G}
&& \frac{{\rm d}}{{\rm d} t} \left( \frac{1}{M^2 G} \right)
= - \frac{133}{10 \pi} \, ,
\\
\label{eq:beta_wM2G}
&& \frac{{\rm d}}{{\rm d} t} \left(
\frac{\omega}{M^2 G} \right) =
\frac{5}{3 \pi} \left( \omega^2
+ 3 \omega + \frac{1}{2} \right) \, ,
\\
\label{eq:betaG}
&& \frac{{\rm d}}{{\rm d} t} \left(\frac{1}{G} \right)
= \frac{5M^2}{3 \pi} \left( \omega
- \frac{7}{40 \omega} \right) \, ,
\\
\label{eq:betalamG}
&& \frac{{\rm d}}{{\rm d} t} \left(
\frac{2 \Lambda}{G} \right) = 
\frac{M^4}{2 \pi} \left(5 + \frac{1}{\omega^2}
\right)
- \frac{4M^2 \Lambda}{3 \pi} \left(
14 + \frac{1}{\omega}
\right)
\end{eqnarray}
where $t= \ln (\mu/\mu_0)$ and all $rhs$ contain
the leading contribution in $G$ ($M^2 G$ is also taken
to be small), with higher powers coming from 
higher loops being neglected. 
Using Eqs. (\ref{eq:betaM2G} and \ref{eq:beta_wM2G}) we solve
for the running of $\omega$,
\begin{equation}
\label{eq:beta_w}
\frac{{\rm d} \omega}{{\rm d} t} 
= \frac{5 M^2 G}{3 \pi} 
\left(
\omega^2 + \frac{549}{50} \omega + \frac{1}{2} 
\right) 
= \frac{5 M^2 G}{3 \pi}  (\omega + \omega_1)(\omega + \omega_2)
\end{equation}
where $\omega_1= 0.0457$ and $\omega_2 = 10.9343$. 
Eq. (\ref{eq:beta_w}) shows that $-\omega_1$ and $-\omega_2$ 
are two fixed points, with the former
being repulsive and latter attractive under UV evolution or increasing $t$.
As we see from our propagator Eq. (\ref{eq:grav_prop1}) 
only $\omega$ positive can be in the physical domain. 
That is both the fixed points are in the unphysical domain. 
The $rhs$ of Eq. (\ref{eq:beta_w}) is always positive
for $\omega>0$. This means that $\omega$ is a monotonic 
increasing function of $t$ and vice versa. 
Eq. (\ref{eq:betaM2G}) readily allows us to 
express $M^2G$ in terms of $t$ with which we can 
integrate Eq. (\ref{eq:beta_w}) to obtain,
\begin{equation}
\label{eq:tw}
t= T
\Biggl[
1- 
\left( \frac{\omega + \omega_2}{\omega + \omega_1}
\cdot 
\frac{\omega_0 + \omega_1}{\omega_0 + \omega_2} \right)^{\alpha}
\Biggr] \, ,
\end{equation}
where $T=\frac{10 \pi}{133 M_0^2 G_0} $ 
and $\alpha= 399/50(\omega_2 -\omega_1)$, with
subscript $0$ meaning that the coupling parameters are evaluated at $t=0$ or 
$\mu = \mu_0$. 

Eq. (\ref{eq:beta_w}) transforms any 
evolution in $t$ to evolution in $\omega$.
Using Eqs. (\ref{eq:betaG} and \ref{eq:beta_w}) we get,
\begin{equation}
\label{eq:betalogG}
\frac{{\rm d} \ln G}{{\rm d} \omega} 
= - \frac{\omega - \frac{7}{40 \omega} }
{(\omega + \omega_1) (\omega + \omega_2)} \, .
\end{equation}

$G$ gets maximized at 
$\omega = \sqrt{7/40}$, which lies in the physical domain.
For convenience we choose this point to be our 
reference point $\mu_0$ or $t=0$ and integrate
Eq. (\ref{eq:betalogG}) to obtain,
\begin{equation}
\label{eq:Gw}
\frac{G}{G_0} = \frac{\omega_0}{\omega} \cdot 
\left(
\frac{1 + \omega_1/\omega }
{1 + \omega_1/\omega_0 }\right)^{A1}
\left(
\frac{1 + \omega_2/\omega }
{1 + \omega_2/\omega_0 }\right)^{A2}  \, ,
\end{equation}
where $A1=-0.3473$ and $A2= -1.0027$. 
Eqs. (\ref{eq:betalogG} and \ref{eq:Gw}) show that 
for large $\omega$ or $t$, $G \sim 1/\omega$, thereby
vanishing for large $t$ \cite{Donoghue}, while
for small $\omega$, $G \sim \omega^{7/20}$
reaching a peak at $\omega_0=\sqrt{7/40}$.
Similarly, using Eqs. (\ref{eq:beta_wM2G} and \ref{eq:betaG})
we obtain the running of $M^2/\omega$, which along with
with Eq. (\ref{eq:beta_w}) is integrated to give,
\begin{equation}
\label{eq:M2byw}
\frac{M^2}{\omega} = \frac{M_0^2}{\omega_0} 
\left(
\frac{1 + \omega_1/\omega }
{1 + \omega_1/\omega_0 }\right)^{B1} 
\left(
\frac{1 + \omega_2/\omega }
{1 + \omega_2/\omega_0 }\right)^{B2} \, ,
\end{equation}
where $B1=1.0802$ and $B2= 0.2698$.
We can now analyze $M^2/\mu^2$ from Eq. (\ref{eq:M2byw}).
Alternatively, it is instructive to note that from
Eqs. (\ref{eq:betaM2G}, \ref{eq:betaG} and \ref{eq:beta_w}) we obtain
the running of $\ln (M^2/\mu^2)$,
\begin{equation}
\label{eq:beta_M2mu}
\frac{{\rm d}}{{\rm d} \omega} \ln \left( \frac{M^2}{\mu^2} \right)
= \frac{ \left(
\omega + \frac{399}{50} - \frac{7}{40 \omega}
- \frac{6 \pi}{5 M^2 G}
\right) }
{ (\omega + \omega_1)
(\omega + \omega_2)}
\, .
\end{equation}
This shows that $M^2/\mu^2$ reaches a minima for 
$\omega = \omega_*$ given by 
$
\left(
\omega_* + \frac{399}{50} - \frac{7}{40 \omega_*}
\right)
= \frac{6 \pi}{5 M_*^2 G_*}.
$
Hence by demanding
\begin{equation}
\label{eq:mincond}
\frac{M^2_*}{\mu^2_*} = \frac{6 \pi}{5 \mu^2_* G_*}
\frac{1}{\omega_* + \frac{399}{50} - \frac{7}{40 \omega_*}}
> 1
\end{equation}

we make the $M$-mode not realizable
in the physical GR sector of the theory.
The inequality is easily achieved by choosing $\mu^2_* G_*$ 
appropriately. Perturbative loop expansion requires that 
$M^2G$ is small. Therefore $M$ is a sub-Planckian mass,
yet the running mass as dictated by interactions makes it
physically not realizable even in post Planckian regime.

Analogous to Eq. (\ref{eq:M2byw}) from Eqs. (\ref{eq:beta_wM2G}, \ref{eq:betaG}
and \ref{eq:beta_w}) we obtain,
\begin{equation}
\label{eq:riccion_min}
\frac{{\rm d}}{{\rm d} \omega} 
\ln \left(\frac{M^2}{\omega \mu^2} \right)
= - \frac{3 + \frac{27}{40 \omega} + \frac{6 \pi}{5 M^2 G} }
{(\omega + \omega_1)(\omega + \omega_2)} \, ,
\end{equation}
showing that the Riccion mass relative to $\mu$
decreases monotonically. By a suitable choice we can make the
Riccion to be physically realizable or not. So we conclude that 
there exists unitary physical subspace only 
with the gravitons or along with Riccions.

We find that the solution of $\Lambda$ in the RG flows 
is dictated by $M^2$ and $G$. From Eq. (\ref{eq:betalamG})
because of $M^4$ term $\Lambda$ increase in the UV region.
These terms can be explicitly eliminated by adding fermion
fields to our action. For example adding two spin-$\frac{1}{2}$ Dirac fields
with mass $(5/4)^{1/4}M$ and $M/\sqrt{2\omega}$ \cite{Gorbar}
(these additional fermionic fields can be interpreted as ghosts
normalizing the functional integral).
Consequently the other coefficients 
in Eqs. (\ref{eq:betaM2G}, \ref{eq:beta_wM2G} and \ref{eq:betaG}) do change
in $rhs$ making $\omega_0$, $\omega_1$, $\omega_2$ and $\omega_*$
shift slightly but all conclusions remains unaltered. Furthermore,
in this modified theory the cosmological constant can be 
made to vanish for large $t$.

We  now look at the generality of the results. 
Adding matter fields to our action $S$ modifies the RG flow
Eqs. (\ref{eq:betaM2G}, \ref{eq:beta_wM2G}, \ref{eq:betaG} and 
\ref{eq:betalamG}). In particular in $rhs$ we will have terms which are 
dependent on mass $m$ and couplings defined within the matter theory
but independent of $M$ and $\omega$. For example to one-loop in 
matter fields these have been computed \cite{Gorbar, Larsen, avramidibook}. 
Assuming $m/M<1$,
where $m$ is a typical matter field mass, we notice that $\omega_0$,
$\omega_1$, $\omega_2$, and $\omega_*$ all shift by a small amount
and all our conclusion remain unaltered. We remark that in the absence of
higher derivative terms {\it i.e.} the $R^2$ and conformal term, running of $G$ 
is dominated only by matter fields and $G$ increases in the UV 
regime \cite{Larsen}. In the renormalizable theory given by $S$, the Riccion and $M$-mode 
reverse this tendency and make it vanish.

\begin{figure}
\centerline{
\begin{minipage}[t]{3.5in}
\vspace{0pt}
\centering
\includegraphics[width=3.5in]{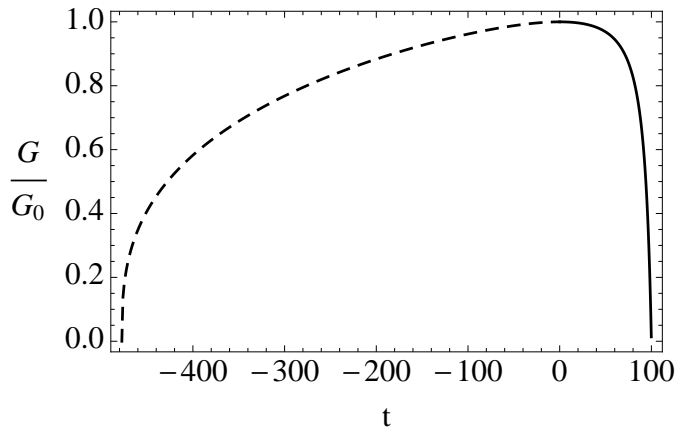}
\end{minipage}
}
 \caption[]{
Running Newton's gravitational coupling.
  }
\label{fig:Gt}
\vspace{-5mm}
\end{figure}

All our calculations are done in the Landau gauge. These calculations
can in principal be done in other gauges, which have additional 
gauge parameters \cite{avramidi, avramidibook}. 
They do not affect Eqs. (\ref{eq:betaM2G}, \ref{eq:beta_wM2G})
and in Eq. (\ref{eq:betaG}) there are additional terms in $rhs$ proportional 
to $M^2$ and $M^2/\omega$, but not $M^2 \omega$. 
Consequently $\omega_1$ and $\omega_2$ are unaltered
but $\omega_0$ shifts. It should also be noted 
that in all these gauge parameter choices, 
for large $\omega$, $G$ still vanishes as $1/\omega$, 
while for small $\omega$, $G$ does vanish
with a different power (as opposed to $\omega^{7/20}$ 
in the Landau gauge).

The mass of the Riccion $M/\sqrt{\omega}$ is seen to be 
gauge parameter dependent, however in all these cases 
it has positive norm. 
We also looked at the unitarity issue in Prentki gauge 
($\partial_i h_{i\mu}=0$, where $i=1,2,3$) \cite{veltman}.
In this gauge the Faddeev-Popov ghost is not propagating and we have 
exactly gravitons, Riccion with positive norm and mass $M/\sqrt{\omega}$
and spin-2 $M$-mode with negative norm and mass $M$, just as in Landau gauge.
In addition there is a vector particle with positive norm and mass $M$ (same as
$M$-mode, RG flow equations have not been computed in this gauge).
We see that in all these gauges the physical GR sector scattering 
matrix is unitary.

Our RG analysis is at $G=\Lambda=0$. However it is should be
noted that higher derivative gravity action Eq. (\ref{eq:hdgact}) was
also considered at a non-trivial fixed point where both $G$ and $\Lambda$
are non-vanishing \cite{Codello, Niedermaier1}. At this
fixed point unitarity issue was considered \cite{Niedermaier1, Niedermaier2}
at one-loop in a generalized de-Donder gauge using 
non-minimal regularization scheme. By doing Wilson's RG analysis 
on Euclidean functional integral, it is shown that the fixed point
enjoys necessary spectral positivity properties over a finite 
range of parameter $\omega$. 

We conclude that for large $\omega$, $G \sim 1/\omega$ is a 
gauge invariant result. 
In the UV regime $1/M^2G$ is the coefficient of conformal
term in the action and this is finite while $R$ and $R^2$ coefficients are 
becoming large as $\omega$.
$R^2$ being always positive makes functional integral better 
behaved as well. In addition the renormalized action enforces
$R$ not to fluctuate away from zero. In infrared the EH term
$R$ naturally is dominant along with cosmological 
constant. In infrared if $\Lambda=0$ (by tuning) the 
vanishing of $G$ again enforces that $R$ fluctuate 
around zero.

\begin{figure}
\centerline{
\begin{minipage}[t]{3.5in}
\vspace{0pt}
\centering
\includegraphics[width=3.5in]{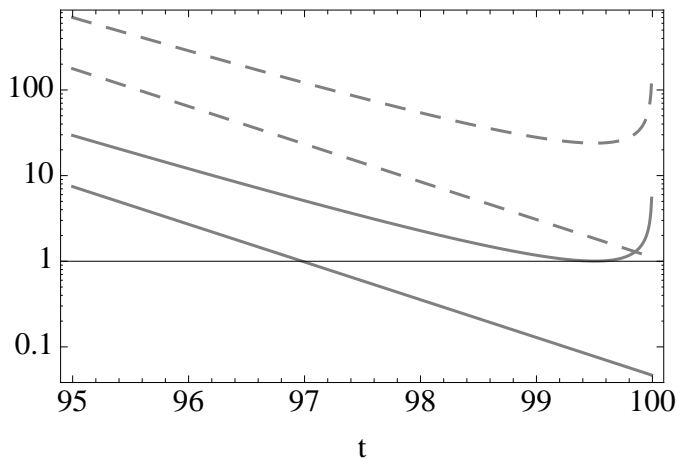}
\end{minipage}
}
 \caption[]{
$M$-mode ratio $M/\mu$ (top) and Riccion mass ratio $M/\mu \sqrt{\omega}$ (bottom).
Dashed lines for $M_*/\mu_*=25$ and solid lines for $M_*/\mu_*=1.05$.
}
\label{fig:crossover}
\vspace{-5mm}
\end{figure}

\section{Discussion}
\label{discuss}

The physical domain $0 \leq \omega <\infty $
translates to a finite range in $t$, namely
\begin{equation}
\label{eq:tmax_min}
\frac{t_{min}}{T} \equiv 
1 - \left(\frac{\omega_2}{\omega_1} 
\frac{\omega_0 + \omega_1}
{\omega_0 + \omega_2}
\right)^{\alpha}
\leq \frac{t}{T}
< 
1 - \left(
\frac{\omega_0 + \omega_1}
{\omega_0 + \omega_2}
\right)^{\alpha}
\equiv \frac{t_{max}}{T}
\end{equation}
From one-loop analysis we see that
there is no solution for $\omega$ from Eq. (\ref{eq:tw}) if 
$t>t_{max}$. For $t< t_{min}$, $\omega$ becomes negative {\it i.e.}
$M^2/\omega$ the Riccion mass square becomes negative
signaling the instability of the vacuum. This infrared issue
needs to be studied, namely the effects of the cosmological 
constant. To the leading order in Feynman loop expansion, 
at both extremes $G$ vanishes as $1/\omega$ for $t \to t_{max}$
or $\omega^{7/20}$ for $t \to t_{min}$.

Including inflationary scenarios the entire cosmological phenomenon
ranges from as early as $10^{-42}$s to the current age 
of the universe $10^{18}$s, a span of about $130$ e-folds.
In the higher derivative 
gravity theory we have essentially the following 
dimensionful parameters $G_0$, $M_0$ and $\mu_0$
($\omega_0$ is set by the maximum of $G$).
These can be traded for one dimensionful parameter $G_0$
and two dimensionless parameters $t_{max}$ and 
$M_*/\mu_*$. By choosing $t_{max}$ large enough for example 
about 100, we plotted $G/G_0$ in Fig. \ref{fig:Gt} and $M/\mu$ 
and $M/\sqrt{\omega} \mu$ in Fig. \ref{fig:crossover}. 
For $t_{max}=100$, if $M_*/\mu_* > 22.52$
then both $M$-mode and Riccion are not in 
physical GR sector. In the Fig. \ref{fig:crossover}, the 
dashed line plot represent such a case for 
$M_*/\mu_* =25$. If $1< M_*/\mu_* \leq 22.52$,
then Riccion is realized. In the Fig. \ref{fig:crossover}, the solid
lines represent an extreme case for $M_*/\mu_* =1.05$.
From Fig. \ref{fig:crossover} we notice that Riccion is physically 
realizable in high energy scattering processes 
at most for about three e-folds before $t_{max}=100$. 

From Fig. \ref{fig:Gt} we note that the entire cosmology may span
around the maximum of $G$ accounting for sub-Planckian regime. 
Eventually in the post-Planckian regime $>\, 10^{19}$ GeV ($10^{-43}$s),
the gravitation coupling goes to zero at some finite energy
$\mu_0 e^{t_{max}}$ (from Eq. (\ref{eq:betaG}) for $\omega> \sqrt{7/40}$
and $M^2G$ positive, $G$ exponentially reaches zero and remains 
at zero). Naturally the question of physics
beyond this scale is raised. 

To discuss the range of validity of this analysis,
we rewrite the propagator Eq. (\ref{eq:grav_prop1}) 
\begin{equation}
\label{eq:grav_prop2}
D_{\mu\nu, \alpha\beta} =
 \frac{i \, 16 \pi}{(2 \pi)^4} \cdot 
\Biggl[
\frac{M^2G}{\omega} \frac{(P_s)_{\mu\nu, \alpha\beta}}{q^2 - M^2/\omega + i \epsilon}
- M^2 G  \frac{2 \, (P_2)_{\mu\nu, \alpha\beta}}{q^2 - M^2+ i \epsilon}
\Biggr] \cdot \frac{1}{q^2 + i\epsilon} 
\end{equation}
From Eq. (\ref{eq:hdgact})  and Eq. (\ref{eq:grav_prop2}) we 
see that there are three perturbative parameters $G$, $\frac{M^2G}{\omega}$
and $M^2G$ and in the UV limit the important dimensionless 
perturbative parameters are latter two. 
Under these circumstances we find $M^2G/\omega$ and $G$
both vanish at a finite energy scale $\mu_0 e^{t_{max}}$.
These results are valid if $M^2G$ is small as shown in the figures.
However from Eq. (\ref{eq:betaM2G}) we also know that eventually
$M^2G$ does become large before it reaches Landau singularity.
We envisage two possible alternatives. Higher loop 
corrections may introduce terms like $(M^2G) \ln \omega$ etc, such
that the Landau singularity is avoided and $t_{max}$ approach 
infinity. Another alternative could be $M^2G$ becomes of order 
one, consequently we need to consider the conformal
term in Eq. (\ref{eq:hdgact}) non-perturbatively. 

Finally we conclude that the action $S$, Eq. (\ref{eq:hdgact}) 
describes a perturbative quantum 
gravity as self consistent, renormalizable and unitary theory 
of gravitons and the curvature cannot become
singular, in particular it cannot fluctuate wildly at sub-Planckian 
\cite{Starobinsky} or post Planckian regimes
consistent with known cosmology. 
Its predictions asymptotically beyond Planck scale needs 
to be investigated further.



\end{document}